\documentclass[aps,amssymb,showpacs,twocolumn]{revtex4}
\usepackage{amssymb}
\usepackage{graphicx}
\usepackage{color}
\usepackage{textcomp}
\usepackage{ulem}
\begin{document}

\title{Quark anomalous magnetic moment leads to the inverse magnetic catalysis phenomena of chiral restoration and deconfinement phase transitions in $\mu_B-T$ plane}
\author{Shijun Mao}%
 \email{maoshijun@mail.xjtu.edu.cn}
\affiliation{School of Physics, Xi'an Jiaotong University, Xi'an, Shaanxi 710049, China}

\begin{abstract}
The effect of quark anomalous magnetic moment (AMM) to chiral restoration and deconfinement phase transitions in baryon chemical potential-temperature $(\mu_B-T)$ plane under magnetic fields is investigated in frame of a Pauli-Villars regularized PNJL model. It's found that the quark AMM plays the role of inverse catalysis to the phase transitions, and large quark AMM will change the magnetic catalysis phenomena of phase transitions to inverse magnetic catalysis in the whole $\mu_B-T$ plane. For a fixed magnetic field, the critical temperature $T_c$ and critical baryon chemical potential $\mu_B^c$ decreases with quark AMM. The stronger the magnetic field is, the inverse catalysis effect of AMM becomes more important. For a small AMM $\kappa=\kappa_1$, it shows the magnetic catalysis effect for critical temperature $T_c$ at vanishing $\mu_B$ with increasing magnetic field, and (inverse) magnetic catalysis effect for critical baryon chemical potential $\mu_B^c$ at vanishing $T$ under (weak) strong magnetic field. At finite $T$ and $\mu_B$, there exist some crossings of the phase transition lines with different magnetic field. For a large AMM $\kappa=\kappa_2$, we obtain the inverse magnetic catalysis effect in the whole $\mu_B-T$ plane, and no crossings of phase transition lines happen.
\end{abstract}

\date{\today}
\maketitle

\section{introduction}
The study on QCD phase structure is recently extended to including external electromagnetic fields, motivated by the strong magnetic field in the core of compact stars and in the initial stage of relativistic heavy ion collisions~\cite{review0,review1,review2,review3,review4,review5}. From recent lattice QCD simulations with a physical pion mass, while the chiral condensate is enhanced in vacuum, the critical temperature of the chiral restoration phase transition drops down with increasing magnetic field, which is the inverse magnetic catalysis effect~\cite{lattice1,lattice11,lattice2,lattice3,lattice4,lattice5}. On the other hand, lattice simulations on the Polyakov loop also support the inverse magnetic catalysis for deconfinement phase transition, with a decreasing critical temperature as the magnetic field grows~\cite{lattice4,lattice5}. At finite baryon chemical potential, Lattice QCD has the sign problem.

On analytical side, in the presence of a uniform external magnetic field ${\bold B} =B{\it {{\bold e}_z}}$, the energy dispersion of charged fermions takes the form $E=\sqrt{p_z^2+2eBl+m^2}$ with the momentum $p_z$ along the direction of magnetic field and the Landau level $l=0,1,2,...$~\cite{landau}. Due to this fermion dimension reduction, almost all model calculations at mean field level report the magnetic catalysis effect both in vacuum (increasing quark mass with magnetic field) and at finite temperature (increasing critical temperature for chiral restoration and deconfinement phase transitions), see review~\cite{review0,review2,review4,review5} and the references therein. How to explain the inverse magnetic catalysis phenomena is an open question~\cite{fukushima,mao,kamikado,bf1,bf12,bf13,bf2,bf3,bf4,bf5,bf51,bf52,bf8,bf9,bf11,db1,db2,db3,db5,db6,pnjl1,pnjl2,pnjl3,pnjl4,pqm,ferr1,ferr2,mhuang}. For critical baryon chemical potential, people find the inverse magnetic catalysis effect at weak magnetic field, decreasing with magnetic field and accompanied with oscillations, and magnetic catalysis effect at strong magnetic field, increasing with magnetic field, see review~\cite{review1} and the references therein.

The magnetic field also affects the radiative corrections of the fermion self-energy, which corresponds to the coupling between the magnetic field and the fermion anomalous magnetic moment (AMM)~\cite{amme1,amme2,amme3,amm0,amm1,amm11,amm2,amm3,amm4,amm5}. This gives rise to a new term $\frac{1}{2}\kappa_f Q_f \sigma_{\mu \nu} F^{\mu \nu}$ in the Dirac Hamiltonian, with field tensor $F^{\mu \nu}$, spin tensor $\sigma_{\mu \nu}=\frac{i}{2}[\gamma_\mu, \gamma_\nu]$ and quark electrical charge $Q_f$, and the coefficient $\kappa_f$ is identified as the fermion AMM. The AMM term in the Hamiltonian changes the energy spectrum of fermions by removing the spin degeneracy and affects the properties of magnetized systems~\cite{ferr1,ferr2,mhuang,amm5,amm6,amm7,ammp1,ammp2,meijie}. In the lowest Landau level approximation, the quark AMM $\kappa_f$ reduces the effective quark mass with $m_{\text {eff}}=m- \kappa_f |Q_f B|$, and plays the role of inverse catalysis effect in the chiral restoration and deconfinement phase transitions~\cite{meijie,PLB722}.

Including the AMM term, the chiral restoration and deconfinement phase transitions under external magnetic field are determined by the two competing factors, the catalysis effect of magnetic field to quark mass and inverse catalysis effect of quark anomalous magnetic moment. It's obtained that the critical temperature at vanishing baryon chemical potential increases with magnetic field for a small quark AMM, but decreases with magnetic field for a large quark AMM~\cite{amm6,amm7,meijie}. In this paper, we will study the quark AMM effect on the phase structure of chiral restoration and deconfinement in baryon chemical potential-temperature $(\mu_B-T)$ plane under external magnetic field. As an extension of our previous work~\cite{meijie} at finite temperature and vanishing baryon chemical potential, we make use of the Polyakov-extended Nambu--Jona-Lasinio (PNJL) model~\cite{pnjl5,pnjl6,pnjl7,pnjl8,pnjl9,pnjl10,pnjl12}. It is found that with large enough quark AMM, the inverse magnetic catalysis phenomena will be observed in the whole $\mu_B-T$ plane.

After the introduction, we extend our PNJL framework with AMM to finite temperature and finite baryon chemical potential case in Sec.\ref{model}. The physical results and discussions are presented in Sec.\ref{result}, and Sec.\ref{sum} is a brief summary.

\section{The model}
\label{model}
The two-flavor PNJL model under external electromagnetic field is defined through the Lagrangian density~\cite{pnjl5,pnjl6,pnjl7,pnjl8,pnjl9,pnjl10,pnjl12} in chiral limit,
\begin{eqnarray}
\mathcal{L}&=&\bar{\psi}(x)\left(i\gamma^{\mu}D_{\mu}+\frac{1}{2} a {\sigma}^{\mu\nu} F_{\mu\nu}\right)\psi(x)\\
&&+\frac{G}{2}\left\{[\bar{\psi}(x)\psi(x)]^2+[\bar{\psi}(x)i\gamma_5 \vec{\tau}\psi(x)]^2 \right\}-\mathcal{U}(\Phi,\bar{\Phi}).\nonumber
\label{lagrangian}
\end{eqnarray}
For the chiral section in the Lagrangian, the covariant derivative $D^\mu=\partial^\mu+i Q A^\mu-i {\cal A}^\mu$ couples quarks to the two external fields, the magnetic field ${\bf B}=\nabla\times{\bf A}$ and the temporal gluon field  ${\cal A}^\mu=\delta^\mu_0 {\cal A}^0$ with ${\cal A}^0=g{\cal A}^0_a \lambda_a/2=-i{\cal A}_4$ in Euclidean space. The gauge coupling $g$ is combined with the SU(3) gauge field ${\cal A}^0_a(x)$ to define ${\cal A}^\mu(x)$, $\lambda_a$ are the Gell-Mann matrices in color space, and $Q=diag(Q_u, Q_d)=diag(2e/3,-e/3)$ is the quark charge matrix in flavor space. The quark anomalous magnetic moment (AMM) is introduced by the term $\frac{1}{2} a {\sigma}_{\mu\nu} F^{\mu\nu}$, with spin tensor $\sigma_{\mu\nu}=\frac{i}{2}\left[\gamma_{\mu},\gamma_{\nu}\right]$, the Abel field strength tensor $F_{\mu\nu}=\partial_{[\mu,}A_{\nu]}$, and the quark AMM ${a}=Q { \kappa}$ and ${ \kappa}=diag({ \kappa}_u, { \kappa}_d)$ in flavor space. $G$ is the coupling constant in the scalar and pseudo-scalar channels. The order parameter to describe chiral restoration phase transition is the chiral condensate $\langle\bar\psi\psi\rangle$ or the dynamical quark mass $m=-G \langle \bar\psi\psi\rangle$. The Polyakov potential ${\cal U}(\Phi,{\bar \Phi})$ is related to the Z(3) center symmetry and simulates the deconfinement~\cite{pnjl6}
\begin{eqnarray}
\frac{\mathcal{U}}{T^{4}}=-\frac{b_{2}(t)}{2}\bar{\Phi}\Phi-\frac{b_3}{6}\left(\bar{\Phi}^3+\Phi^3\right)+\frac{b_4}{4}\left(\bar{\Phi}\Phi\right)^2,
\end{eqnarray}
where the Polyakov loop is defined as $\Phi=\left({\text {Tr}}_c L \right)/N_c$ and ${\bar \Phi}=\left({\text {Tr}}_c L^\dagger \right)/N_c$, with $L({\bf x})={\cal P} \text {exp}[i \int^{1/T}_0 d \tau {\cal A}_4({\bf x},\tau)]= \text {exp}[i {\cal A}_4/T ]$. The coefficient $b_2(t)=a_0+a_1 t+a_2 t^2+a_3 t^3$ with $t=T_0/T$ is temperature dependent, and the other coefficients $b_3$ and $b_4$ are constants. Polyakov loop is considered as the order parameter to describe the deconfinement phase transition~\cite{pnjl5,pnjl6,pnjl7,pnjl8,pnjl9,pnjl10,pnjl12}. To simplify calculations, we assume a constant magnetic field ${\bf B}=(0, 0, B)$ along the $z$-axis and constant quark AMM $\kappa$ (or $a$).

Taking mean field approximation, the thermodynamic potential at finite temperature $T$ and baryon chemical potential $\mu_B$ contains the mean field part and quark part
\begin{eqnarray}
\label{omega1}
\Omega_{\text {mf}} &=&{\cal U}+ \frac{m^2}{2 G}+\Omega_q,\\
\Omega_q &=& - \sum_{f,n,s} \int \frac{d p_z}{2\pi} \frac{|Q_f B|}{2\pi} \nonumber \\&&\ \ \ \left[3E_f+ T\ln\left(1+g^-\right)+ T\ln\left(1+g^+\right)\right],\nonumber
\end{eqnarray}
where
\begin{eqnarray}
g^-&=&3\Phi e^{- E_f^-/T}+3{ \bar{\Phi}}e^{-2 E_f^-/T}+e^{-3 E_f^-/T},\nonumber\\
g^+&=&3 {\bar{\Phi}} e^{- E_f^+/T}+3{ \Phi}e^{-2 E_f^+/T}+e^{-3 E_f^+/T},\nonumber
\end{eqnarray}
with quark energy $E_f^\pm=E_f \pm \frac{\mu_B}{3}$, and $E_f=\sqrt{p^2_z+\left(\sqrt{\left(2 n+1-s \xi_f \right) |Q_f B|+m^2}-s \kappa_f Q_f B \right)^2 }$ for flavor $f$, longitudinal momentum $p_z$, Landau level $n$, spin $s$ and sign factor $\xi_f={\text {sgn}}(Q_f B)$.

The ground state is determined by minimizing the thermodynamic potential, $\partial\Omega_{\text {mf}}/\partial m=0$, $\partial\Omega_{\text {mf}}/\partial {\bar {\Phi}}=0$ and $\partial\Omega_{\text {mf}}/\partial \Phi=0$, which leads to the three coupled gap equations for the three order parameters $m, \Phi, {\bar {\Phi}}$,
\begin{eqnarray}
\label{gap1}
&& m\left( \frac{1}{2G}+\frac{\partial \Omega_q}{\partial m^2}\right)=0,\\
\label{gap2}
&& \frac{\partial {\cal U}}{\partial \Phi}+\frac{\partial \Omega_q}{\partial \Phi}=0,\\
\label{gap3}
&& \frac{\partial {\cal U}}{\partial {\bar {\Phi}}}+\frac{\partial \Omega_q}{\partial {\bar {\Phi}}}=0.
\end{eqnarray}
From Eq.(\ref{gap1}), we can always find a solution $m=0$. The chiral restoration phase transition happens when non-vanishing quark mass $m$ turns to zero. At this time, the coupled gap equations Eq.(\ref{gap1}) and Eq.(\ref{gap2}),(\ref{gap3}) become decoupled. In the chiral restoration phase with $m=0$, we only need to solve Polyakov loop $\Phi$ and $\bar {\Phi}$ from Eq.(\ref{gap2}) and Eq.(\ref{gap3}). Thus, we obtain the same critical temperature for chiral restoration and deconfinement phase transitions in chiral limit~\cite{mao,meijie}.


Because of the contact interaction among quarks, NJL models are nonrenormalizable, and it is necessary to introduce a regularization scheme to remove the ultraviolet divergence in momentum integrations. The magnetic field does not cause extra ultraviolet divergence but introduces discrete Landau levels and anisotropy in momentum space. The usually used hard/soft momentum cutoff regularization schemes do not work well in magnetic field, since the momentum cutoff together with the discrete Landau levels will cause some nonphysical results~\cite{amm2,amm6,amm7,reg1,reg2,reg3,reg4,reg6,reg7}, such as the oscillations of chiral condensate and critical temperature, tachyonic pion mass, and the breaking of law of causality for Goldstone mode. In this work, we take into account the gauge covariant Pauli-Villars regularization scheme~\cite{mao,meijie}, where the quark momentum runs formally from zero to infinity, and the nonphysical results are cured~\cite{mao,reg4,reg6}. Under the Pauli-Villars scheme, one introduces the regularized quark energy $E_{f}^i=\sqrt{p^2_z+M_{\text {eff}}^2+a^i \Lambda^2}$ with $M_{\text {eff}}=\sqrt{\left(2 n+1-s \xi_f \right) |Q_f B|+m^2}-s \kappa_f Q_f B$ and the summation and integration $\sum_n\int dp_z/(2\pi)F(E_f)$ is replaced by $\sum_n\int dp_z/(2\pi)\sum_{i=0}^N c^iF(E_f^i)$. In our numerical calculations, all the Landau levels and longitudinal momenta are considered, and the regularization is applied to all the summations and integrations in Eqs(\ref{omega1})-(\ref{gap3}).

In chiral limit there are two parameters, the quark coupling constant $G$ and Pauli-Villars mass parameter $\Lambda$. By fitting the pion decay constant $f_\pi=93$ MeV and chiral condensate $\langle \bar \psi \psi \rangle =(-250\ \text{MeV})^3$ in vacuum, the two parameters are fixed to be $G=7.04$ GeV$^{-2}$ and $\Lambda=1127$ MeV in Pauli-Villars scheme with number of regulated quark masses $N=3$, and coefficients $a^1=1, c^1=-3$, $a^2=2, c^2=3$, $a^3=3, c^3=-1$, which are determined by constraints $a^0=0$, $c^0=1$, and $\sum_{i=0}^N c^i\left(m^2+a^i\Lambda^2\right)^{L}=0$ for $L=0,1,\cdots N-1$~\cite{meijie}. For the Polyakov potential, its temperature dependence is from the lattice simulation, and the parameters are chosen as $a_0=6.75$, $a_1=-1.95$, $a_2=2.625$, $a_3=-7.44$, $b_3=0.75$, $b_4=7.5$ and $T_0=270$ MeV~\cite{pnjl6}. To evaluate the effect of quark AMM, we consider two typical sets of parameters $\kappa$, $\kappa_u^{(1)}=0.00995\ {\text {GeV}}^{-1},\ \kappa_d^{(1)}=0.07975\ {\text {GeV}}^{-1}$ and $\kappa_u^{(2)}=0.29016\ {\text {GeV}}^{-1},\ \kappa_d^{(2)}=0.35986\ {\text {GeV}}^{-1}$, which are phenomenologically determined by fitting the nucleon magnetic moments~\cite{amm2,amm6,amm7,meijie,fit56}. The results with vanishing AMM $\kappa_0=0$ are listed as comparisons.

\section{results}
\label{result}
\begin{figure}[hb]
\includegraphics[width=7cm]{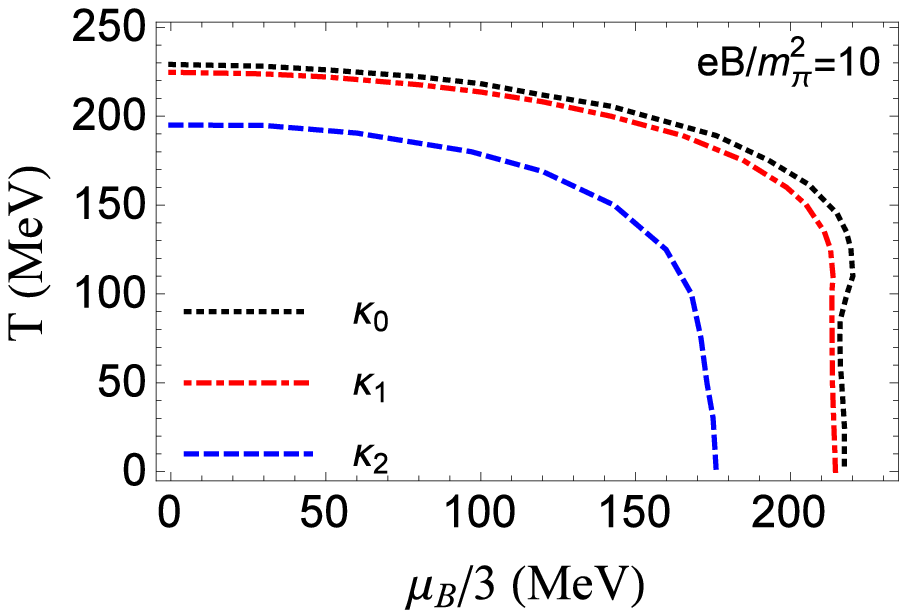}
\includegraphics[width=7cm]{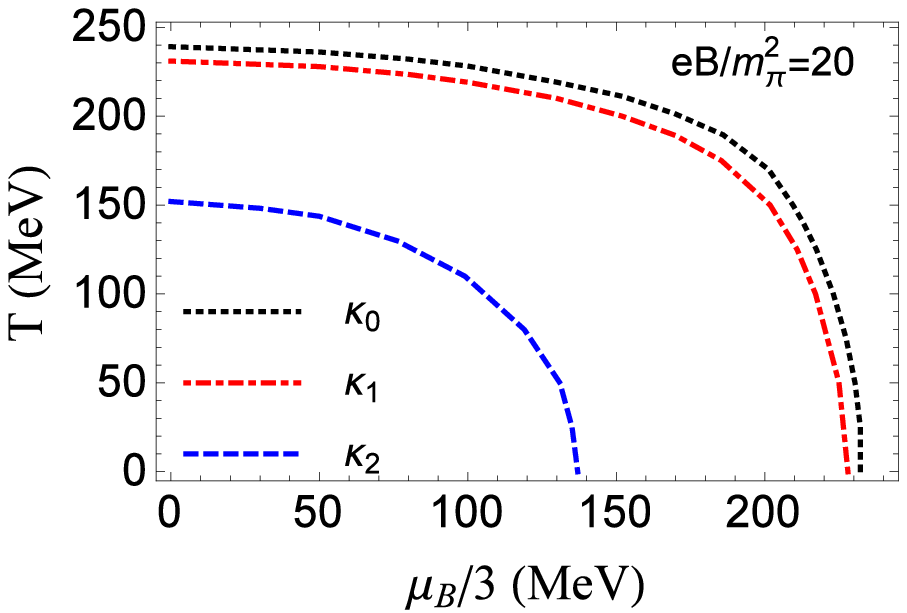}
\caption{The phase diagram with different AMM $\kappa=\kappa_0,\kappa_1,\kappa_2$ in baryon chemical potential and temperature $(\mu_B-T)$ plane at finite magnetic field $eB/m^2_{\pi}=10$ (upper panel) and $eB/m^2_{\pi}=20$ (lower panel)}
\label{fig2}
\end{figure}

In the lowest-Landau-level approximation, we have quark energy $E_f=\sqrt{p^2_z+m_{\text {eff}}^2 }$, with effective quark mass $m_{\text {eff}}=m- \kappa_f |Q_f B|$. The quark AMM affects the system through the contribution to the effective quark mass. For positive $\kappa_f$, the effective quark mass $m_{\text {eff}}$ will be smaller than quark mass $m$, which indicates that the quark AMM induces an inverse catalysis effect to the phase transitions. In Fig.\ref{fig2}, we make comparison of the phase diagram with different AMM $\kappa=\kappa_0,\kappa_1,\kappa_2$ in baryon chemical potential and temperature $(\mu_B-T)$ plane at finite magnetic field $eB/m^2_{\pi}=10, 20\ (m_{\pi}=134\ {\text{MeV}})$. With fixed magnetic field and AMM, the chiral restoration and deconfinement phase transitions happen as increasing temperature and/or baryon chemical potential, which is of first order in the whole $\mu_B-T$ plane at non-vanishing AMM. We obtain lower critical temperature and critical baryon chemical potential with larger AMM at fixed magnetic field, which demonstrates the inverse catalysis effect of quark AMM. Moreover, the inverse catalysis effect of AMM depends on the magnetic field, see the AMM term $\kappa_f |Q_f B|$ in quark energy $E_f$. Without magnetic field, the phase transition lines coincide for different AMM. For stronger magnetic field, this AMM term becomes more important, and hence the separation between phase transition lines with different AMM $\kappa$ becomes further.

\begin{figure}[hb]
\includegraphics[width=7cm]{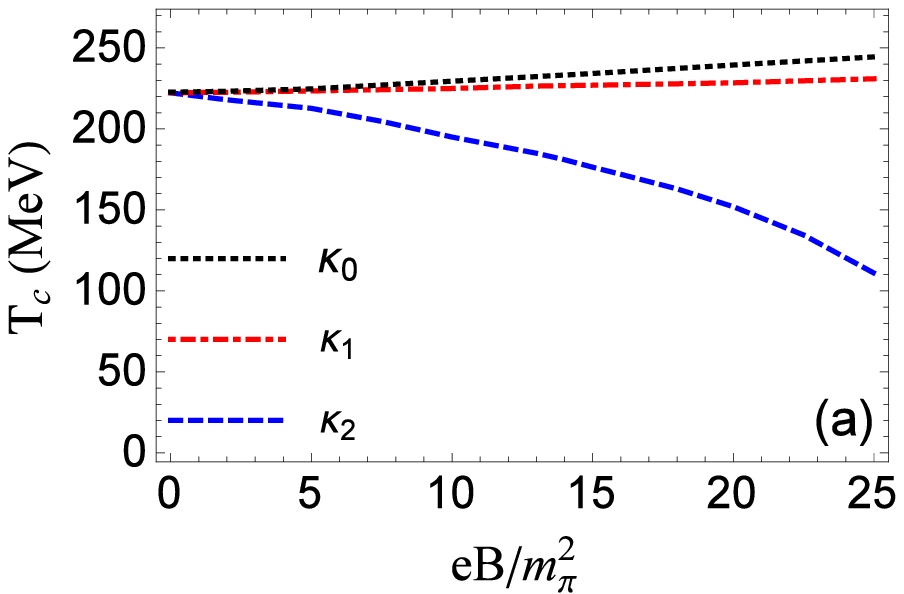}
\includegraphics[width=7cm]{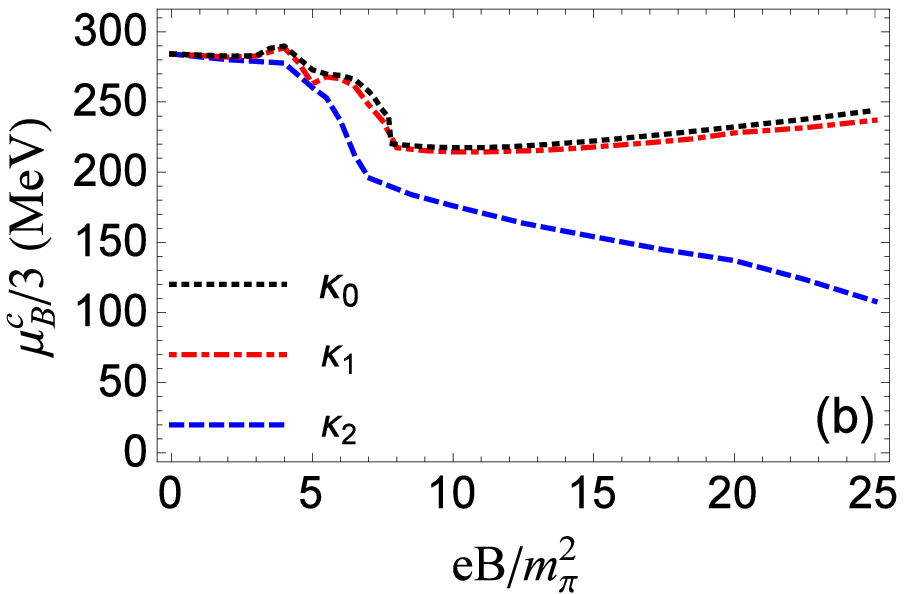}
\caption{The phase diagram of chiral restoration and deconfinement phase transitions in $eB-T$ and $eB-\mu_B$ plane for different quark AMM ${\kappa}$}
\label{fig1}
\end{figure}

With vanishing quark AMM $\kappa_f=0$, the magnetic field causes the catalysis effect to the quark mass~\cite{review0,review2,review4,review5}. Including quark AMM, there appears an inverse catalysis effect caused by the term $\kappa_f |Q_f B|$. Therefore, the chiral restoration and deconfinement phase transitions under external magnetic field are controlled by the competition between the catalysis and inverse catalysis effect. Firstly, we discuss the magnetic catalysis and inverse magnetic catalysis phenomena for critical temperature $T_c$ at vanishing $\mu_B$ and critical baryon chemical potential $\mu_B^c$ at vanishing $T$ in Fig.\ref{fig1}, where the phase transition lines of chiral restoration and deconfinement are depicted with fixed quark AMM $\kappa=\kappa_0,\ \kappa_1,\ \kappa_2$. Fig.\ref{fig1}a is the phase diagram in $eB-T$ plane at vanishing baryon chemical potential~\cite{meijie}. With fixed quark AMM $\kappa=\kappa_0$, the critical temperature $T_c$ increases with magnetic field, that is the magnetic catalysis phenomena. For a small quark AMM $\kappa=\kappa_1$, critical temperature still increases with magnetic field, but with a slower increase ratio. For a large quark AMM $\kappa=\kappa_2$, the critical temperature decreases with magnetic field, that is the inverse magnetic catalysis phenomena. Fig.\ref{fig1}b is the phase diagram in $eB-\mu_B$ plane at vanishing temperature. Without AMM, the critical baryon chemical potential $\mu_B^c$ shows inverse magnetic catalysis at weak magnetic field, decreasing with magnetic field and accompanied with the oscillations, and magnetic catalysis at strong magnetic field, increasing with magnetic field. The turning point is located at $eB/m^2_\pi=10$. For a small quark AMM $\kappa=\kappa_1$, critical baryon chemical potential first decreases and then increases with magnetic field, but the turning point is retarded to $eB/m^2_\pi=11$ and the oscillations are weakened. For a large quark AMM $\kappa=\kappa_2$, the critical baryon chemical potential decreases with magnetic field in the whole region, with very weak oscillations. It is the large quark AMM that changes the magnetic catalysis phenomena for $T_c$ and $\mu_B^c$ to the inverse magnetic catalysis.

\begin{figure}[hb]
\includegraphics[width=7cm]{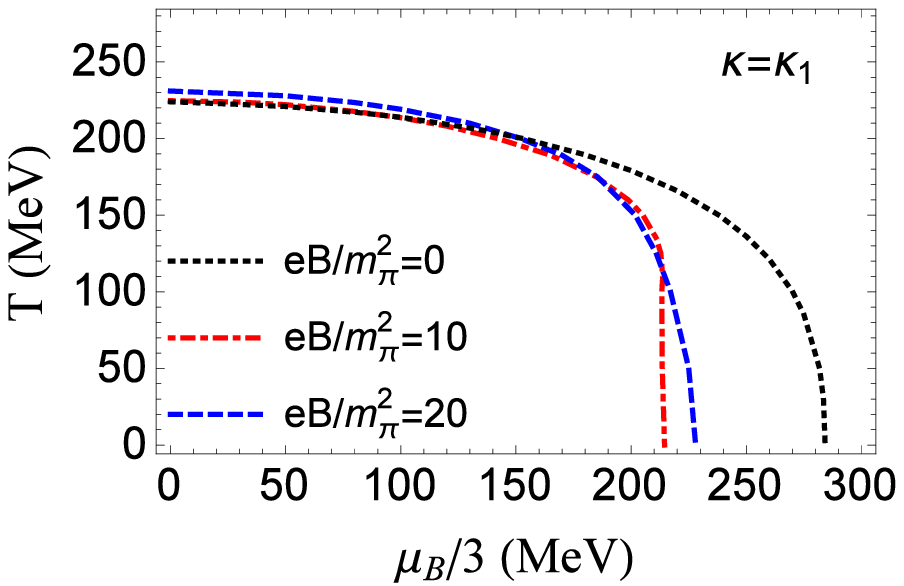}
\includegraphics[width=7cm]{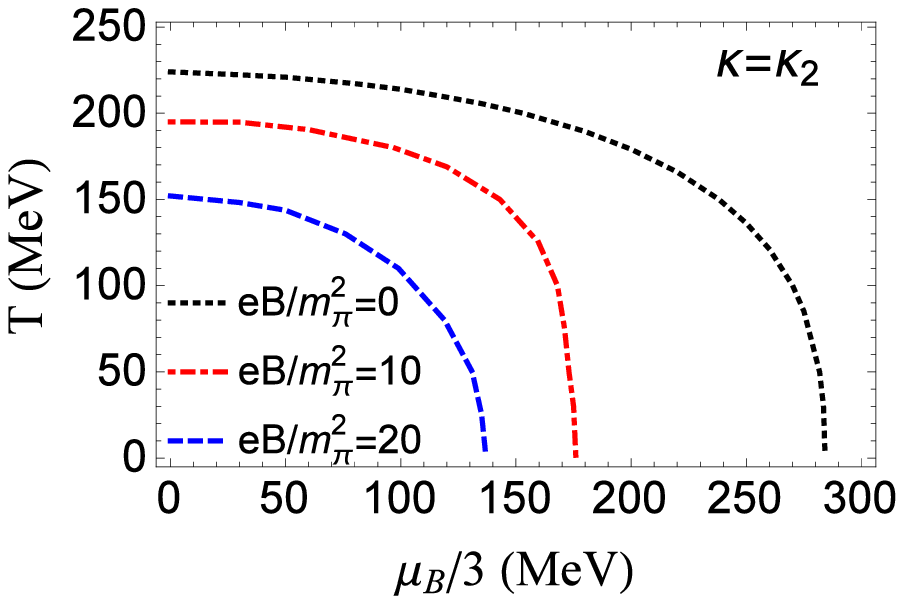}
\caption{The magnetic field effect on the phase structure in $\mu_B-T$ plane with fixed AMM $\kappa=\kappa_1$ (upper panel) and $\kappa=\kappa_2$ (lower panel)}
\label{fig3}
\end{figure}

Fig.\ref{fig3} plots the phase structure at finite magnetic field in $\mu_B-T$ plane with fixed AMM $\kappa=\kappa_1$ (upper panel) and $\kappa=\kappa_2$ (lower panel). With a small AMM $\kappa=\kappa_1$, there exist some crossings for the phase transition lines with different magnetic fields at finite temperature and baryon chemical potential. With finite magnetic field, {\it e.g.} $eB/m^2_\pi=10,20$, we obtain higher $T_c$ but lower $\mu_B^c$ than $eB/m^2_\pi=0$ case. The crossing between phase transition lines with vanishing magnetic field $eB/m^2_\pi=0$ and finite magnetic field $eB/m^2_\pi=10,20$ happens. Comparing finite magnetic field cases with $eB/m^2_\pi=10$ and $eB/m^2_\pi=20$, although we have lower $T_c$ and $\mu_B^c$ at $eB/m^2_\pi=10$, the crossing still happens at finite temperature and baryon chemical potential. With a large AMM $\kappa=\kappa_2$, the inverse magnetic catalysis effect shows in the whole $\mu_B-T$ plane, and thus there is no crossing for the phase transition lines with different magnetic field.

\section{summary}
\label{sum}
The effect of quark anomalous magnetic moment (AMM) to chiral restoration and deconfinement phase transitions in $\mu_B-T$ plane under magnetic fields is investigated in frame of a Pauli-Villars regularized two-flavor PNJL model. Different from the catalysis effect of magnetic field to the quark mass, quark AMM plays the role of inverse catalysis to the phase transitions. The competition between the catalysis and inverse catalysis effect determines the chiral restoration and deconfinement phase transitions under external magnetic field. Large quark AMM will change the magnetic catalysis phenomena in phase transitions to inverse magnetic catalysis in the whole $\mu_B-T$ plane. 

%

\noindent {\bf Acknowledgement:}
The work is supported by the NSFC Grant 11775165.

\end{document}